# Recognition of Cyber-Intrusion patterns in user cognitive behavioural characteristics for remote identification

**Ahmet Orun[1] , Emre Orun[2] and Fatih Kurugollu[3]**

[1]De Montfort University, Faculty of Computing, Engineering and Media. Leicester UK
Email: aorun@dmu.ac.uk.
[2]University of Birmingham, School of Psychology. Birmingham, UK
Email: e.c.orun@bham.ac.uk.
[3]University of Sharjah, Department of Computer Science. Sharjah UAE
Email: fkurugollu@sharjah.ac.ae

## Abstract

Cyber-attacks keep threatening global networks and information infrastructures. The threat is getting more and more destructive and hard to counter day by day as the global networks continue to enlarge exponentially with limited security counter-measures. As this fact requires more sophisticated methods and techniques in urgency, a multidisciplinary remote cognitive observation technique is proposed in this paper to meet today's cybersecurity needs. The proposed method introduces a non-traditional Cognitive Psychology and Artificial Intelligence (AI) based remote threat identification which can be considered during the cyber security system design. It also enables to access the cognitive behavioural parameters of an intruder/hacker remotely without any physical contact via online connection, disregarding the distance of the thread. The ultimate goal of this work is to develop a supplementary cognitive cyber security tool for next generations secure online banking, finance or trade systems.

## 1 Introduction

The series of interactions between a user and computer system connected to a network often exhibit some personal characteristics of cognitive information which is specific to the user by which he/she may be identified and then classified within a certain category in terms of security concerns such as "intruder", "suspicious" or "innocent".

There are many aspects of human cognitive psychology. Some of them are the most effective ones on the motives of cyber-crimes originated from the different factors like radical ideas, criminal tendency or similar negative characteristics. Hence any defence system used against the cyber-crimes cannot be designed distinctly from the well-known psychological factors and related cognitive models. The proposed system in this paper comprises not only such factors and models but also is associated with Artificial Intelligence techniques (e.g. Bayesian network) in a unified form to make a more robust data analysis and a decision making process. Bayesian networks (BN) provides semantic representation of cause-effect relationships between the attributes which increases the speed and accuracy of decision making process. The proposed method unifies the cognitive factors with BN which has been used in identifying system users via cognitive parameters in a serious-game environment and in use of graphical domain (Orun & Seker, 2012) (Ustebay et al., 2014). Within this study, our proposed cyber-intrusion monitoring and defence system relies on three basic functional units, which are detection/diagnosis, prevention and counter-attack where the first and last ones operates in association with the cognitive related components including remote data collection, data analysis and parameter extraction/evaluation. Due to time limitation on such systems' online operations, most techniques and methods have to function in real-time mode like data flow, instant Bayesian classification, cognitive-biometric template matching, prompt analysis and decision making, etc. Such time limitation may also require some operational optimisations (e.g. best parameter subset selection in Orun and Aydin, 2010).

Cognitive psychology proves immensely applicable in practical settings (Hayes, 2000) including cyber security domain (Irigbhogbe, 2015). It is related to human's mental process like attention, memory usage, perception, thinking, problem solving and language use. The human cognition would be in two different forms such as; conscious or unconscious, conceptual or intuitive, abstract or concrete. More often the human's knowledge is used by cognitive process to generate new knowledge, as a field of cognitive psychology this phenomenon has already been inspired by computer sciences. Using computers in collecting cognitive data has always been of interest to researcher in the previous studies. For example, Heoft et al. (2003) proposed a method, known as "Concept Map", to analyse the mental representation by a cognitive data acquisition . This approach is extremely useful due to its ability to provide a visual representation.

In regards to use of cognitive factors in cyber security, the cognitive parameters of an individual always have distinct advantages over the conventional-physical biometric ones (e.g. fingerprint, retina, etc.) such as;

- Cognitive parameters can be utilized for any anonymous individual (user) which have not been recorded or even contacted before by the system.
- Cognitive factors reflect the instant behavioural map of their holder like having a bad intention, even if it looks vice verse or not malicious before. Hence they can be used as time domain variables rather than constants.
- Cognitive parameters, by a deeper psycho-analysis, may give an idea of future behaviour of a user which provides a precaution before any further user-system interaction.
- Those parameters can be acquired "contactless" remotely from a user via online connection, disregarding the distance whatever remote they are.
- Due to phenomenon of social interaction between the individuals, any synchronous changes on the cognitive parameters of a group of individuals may indicate a close interaction between them, and also may give clue how they affect or link to each other (e.g. to detect most effective parameter on their communication)
- Cognitive parameters (hence the cognitive behaviour) are always dynamic and subject to change by a dominant external effects without any awareness of the user. This enable the cyber-security team to perform a counter-cyber attack (e.g. by generating an instant stimuli in a pop-up window) to neutralize them.
- Cognitive based identification is domain/machine/IP independent. This overcomes the problem where the intruder has unauthorised access to someone else's machine or uses unattended one for cyber-attack.

Cognitive process are hypothetical as in fact we cannot physically see what a person is thinking but can deduce or conclude from the evidence based on their behaviours rather than from explicit statements. The question of how a computer work and how we process the information has greatly influence cognitive psychology by developments of computer science and analogies. The computer analogy makes cognitive psychology more interested in how the human brains can store information and also input and output information (Irigbhogbe, 2015) Instead of adopting a scheme that is focused on what a user is doing, it is better to provide a system that explain how the user carry out such a task, this study examine how a user manipulates a GUI and the observed cognitive behaviour is compared to a profile that has been previously generated and evaluation on how effective a system can be in identifying the difference in the users behaviour to understand how it can be used as a means of authentication or defence against an attacker.

## 1.1 Similar earlier works

Earlier studies have also tried to mitigate the cyber security threats by use of AI techniques in various approaches. Such as; Guarascio et al. (2022) introduced a collaborative intrusion detection by use of distributed thread intelligence platform. The other method introduced by Avkurova et al. (2022) uses fuzzy logic to identify intruder by detecting the changes of system parameters by him/her. The metrics utilised different than user's behavioural characteristics. However not much quantitative results are presented for the intruder identification process. One study (Karthika and Maheswari, 2022) has used Bayesian network classifier a machine learning algorithm which is also used in our study. But their intrusion detection is based on system's (normal and malicious) traffic data classification rather than intruder's behavioural characteristics. However they obtains 93.5% classification accuracy which is similar to ours (94%) but with very distinct two classes of normal/malicious, whereas our test users are with very identical characteristics and more difficult to classify. Elbasiony et al. (2013) introduced a similar approach to intrusion detection framework based on anomaly detection. Their anomaly detection procedure is achieved by use of weighted K-means algorithm. But they used the data sets KDD99 (Knowledge discovery in databases) available in the internet domain instead of specifically generated data via user behaviours tracking. The main disadvantage of such readily available data is their parametrical control for a specific experiments and also their transparency (e.g. uncertainty level, reliability, etc.) where the user has to rely on data specification information.

## 1.2 Cognitive psychology and user specific patterns

Cognitive psychology is the area which focuses on the mental processes like; perception, thinking, problem solving, memory, attention and language use. Such action may also have a close link with the contemporary scientific fields like; inference methods, pattern recognition, artificial intelligence, behaviour analysis, etc. The human cognition can also be categorised by different factors. They can be conceptual or intuitive, abstract or concrete, conscious or unconscious. The human cognitive process exploit the knowledge already available and produce new knowledges which is highly complex stage depending on personal options, culture, way of thinking, intelligence, experiences, etc. Graphical perception is another area which may potentially be used to observe individual's cognitive behaviours (Orun and Seker, 2012) for individual specific cognitive data acquisition (Orun et al., 2016). Within this work, as a sub-branch of Cognitive psychology, graphical-visual perception of an individual is exploited who may refer to either cyber attacker or innocent user, depending the series of behaviours exhibited while following the graphical-alphanumerical computer screen domain (e.g. web, command windows, writing tools, etc.) and using the system tools (e.g. keyboard, mouse, touch screen, touchpad, etc.

Contemporary psychology has already introduced many important concepts of Cognitive sciences (Hayes,2000) which may potentially be exploited for Cyber Security remote threat detection.

## 1.3 Cognitive learning factors

Human psychology is a pivotal element concerning cyber security since every individual will vary regarding their online search behaviour. For example, certain individuals may be more vulnerable to phishing attacks and inadvertently installing malware by clicking on dubious links received via email, i.e., risk-takers (Moustafa, Bello, & Maurushat, 2021). Countermeasures for such occurrences can be in the form of working-memory training, to ameliorate behavioural traits associated with inducing security breaches, e.g., impulsivity and procrastination (Rosenbaum, Botdorf, Patrianakos, Steinberg, & Chein, 2017). This may not always be simple however, since certain behaviours fall under habit, and may require considerable conscious effort to counter-condition (Weickert, Joinson, A., & Craggs, 2023). Habits can also be problematic as the unconscious cuing of habitual behaviours can be exploited, for example, if an individual is accustomed to ticking a box to receive emails, a website may take advantage of this common behaviour by setting the function in reverse, whereby individuals who hastily tick boxes out of perseveration may elicit unwanted emails (Norberg, Horne, & Horne, 2007). For these reasons it is important to educate web users on how cyber-attacks can occur to mitigate such risks (Thackray, McAlaney, Dogan, Taylor, & Richardson, 2016; Taylor-Jackson, McAlaney, Foster, Bello, Maurushat, & Dale, 2020).

Cognitive learning is defined as a change in knowledge induced by experience (Mayer, 2011), with the recipience of verbal and visual information being processed by sensory memory, working memory, and long-term memory (Mayer, 2009). Cognitive learning is stated to require three processes – selecting, organising, and integrating information, which abide by the "active processing" principle (Mayer, 2011). Studies within the literature also discuss concepts which can be implemented to assist learning, such as the "segmenting principle", which suggests that allowing the learner to control the pace of information exposure (i.e., receiving information in segments rather than long streams) makes it easier to form connections between verbal and visual stimuli (Mariano, 2014). Cognitive learning has often been facilitated by virtual learning environments (Albus, Vogt, and Seufert, 2021; Annansingh, 2019; De Back, Tinga and Louwerse, 2021). Advantages of virtual environments over real environments have been observed in collaborative learning contexts pertaining to increased motivation and attention (Cho, Yeo, Hwang and Yang, 2022), and can foster greater creativity and confidence (Loureiro and Bettencourt, 2014). It is also worth considering elements such as cognitive load (Sweller, 1988), i.e., the mental resources demanded by one's current task, whereby research suggests that learning can be hindered when cognitive demands exceed working-memory capacity (De Jong, 2010).

One of our previous experiments (Orun, 2014) has proven that every individual who interacts with a virtual environment (like cyber-domain) exhibits a different cognitive learning pattern (*CLP*), whose parameters can be quantised by different measures (e.g., progressive learning index, Chi-square test, etc.). *CLP* may play an important role in user identification as well as making the system unpredictable to the certain category of users, against whom the system has to be protected. Some of the initial cognitive parameters collected via the system input tools (e.g., mouse, keyboards, touch screen, GUI, etc.) are as follows:

- Frequency of action as reaction to graphical stimuli (per/min)
- Max/min idle periods of keyboard/mouse activities (per/min)
- Typing error quantity by keyboard (per/h)
- Process speed of keyboard/mouse activities
- Repetitive attitude of activities

It has proven that the objects play important roles in Cognitive psychology in particular iconic representations, as is stated in psychology literature (Hayes, 2000). The iconic representation is linked to visual imagery to represent objects in memory. This cognitive phenomenon has been exploited in our experiments via graphical domains (Figure 1) as it has the potential to be used for remote connection user's stimuli purposes.

# 2 Methods and materials used

Within the characteristic methodology of proposed work various interdisciplinary techniques and tools have been utilized by optimal integration such as: Bayesian Network utilities (PowerPredictor$^{TM}$ classifier and PowerConstructor$^{TM}$ Bayesian Inference tools) in regards to AI, Macro-Recorder cognitive data collection utility with its continuous operation on a PC, mouse/keyboard components and specific GUI for cognitive stimuli graphical tools.

## 2.1 Data collection and analysis

Data collection process for the user's cognitive parameters has to be done separately for system configuration stage (off-line) and the operation stage (on-line). At the first stage this may be initially carried out by two parts of system units like; user-system interface tools and virtual environments. The first category includes the most conventional tools like keyboard, mouse, touch screen and other GUI components, where the user input signals correspond to user's cognitive behaviours most often specific to his/her mental characteristics. Second type of data collection units refer to virtual environment which may be graphical or non-graphical, by which the user exhibits his/her own personal skills and mental characteristics after each system stimuli activating in a certain time period. Stimuli itself triggers the behavioural actions, hence by measuring behavioural response to stimuli, we can clearly comprehend how a stimuli is processed (Lewandowski and Strohmetz, 2009). The virtual environment can be a video game, graphical skill test, etc. in which all activities of the user are recorded. As already utilized within the experiments, Macro Recorder is a program for reading/recording Keyboard and mouse activities, it's also powerful automation tools for converting macro file to EXE file which can be replayed. This tools was used to record all user interaction with the system using Keyboard and mouse, after a user has complete the given task, the program automatically saved every detail and steps taken by the user in two format, first in an executable form which can be replied to physically observe user behaviour with the machine and the second format is in Text/macro format containing user interaction data, these data are exported to excel sheets for further extraction and analysis. The program also gives the ability to insert custom commands like "goto", "if-then" "repeat $x$ times" etc. For carefully analysis, the software was also used for multiply play back, loop the macro playback and for variable playback speed.

Data analysis stage has to be the crucial part of whole system design and development, which is divided by two basic categories a) attribute extraction from the virtual environment data b) comprehensive analysis of the ultimate data set. Attribute extraction can only be made by an expert staff manually due to its complicated characteristics. But fortunately it will be done only ones for building the cognitive data library. Data analysis will be either on comparative based or absolute numerical values, where the first one has to be preferred due to higher accuracy and robustness. As far as the comparison of different data analysis methods are concerned, our previous experiments showed that Bayesian inference and classification method has an advantage over the Neural networks in terms of graphical representation. Bayesian method yields a satisfactory results if the target class attribute contains no more than two classes, otherwise SVR (support vector regression) method lead to more accurate (Seker et al., 2014) classification. The cognitive behaviour attribute extraction from the users' video data is very important part for the data set generation stage, to achieve this, the direct mouse/keyboard signal readings cannot be used as attributes alone, as they may not provide surplus behavioural optional information specific to the user. The raw signals may be the direct consequence of screen graphical survey questions (stimuli), which do not include any specific personal behaviour exhibited by user, due to this fact, "tracking the motions of mouse cursor/keyboard typing style" may be more appropriate factor to extract behaviour attribute than the raw (time-sequence) mouse/keyboard readings. Some of the mouse cursor/keyboard cognitive attributes could be: hesitation, direct line/curve to target object, circulation, clicking speed, keyboard typing style, etc.

The summary of methodology can be described as: The user's cognitive behavioural data are collected via his/her interaction with the specifically designed graphical user interface domain where each graphical member functions as stimuli to enforce the user to act promptly. Those behavioural data then analysed by AI methods (*Bayesian*) to unveil their cognitive properties and characterise the individual user.

## 2.2 Virtual test environment

To simulate the real world, a virtual test environment has to give a sense of real interaction effects like some of globally available social or commercial platforms (e.g. e-bay, Facebook, stock exchange, etc.). This flexibility can only be achieved by virtual reality facilities whose graphical/non-graphical system parameters can be easily adjusted, modified and produced instantly as required. The virtual test environment will be supported by either specially designed computer/console games and high-speed computer graphics which are used as effective stimuli. As far as the procedure of the tests are concerned, the participant are required to carry out their task individually at different times. The computer was set up with the task running. Participant was instructed to not to carry out additional activity with the computer upon the completion of the running task. The experiment was a Graphical test that require user to manipulate some information display on graphical object or text box. No specific amount of time was set to complete the task, this was to put participate at easy and behave under normal circumstance (Orun, et al. 2016)

The extraction of attribute from the video recording using the naked eye was an interpretation of the video method which is commonly adopted in psychological experiment (Horwitz et al., 2010) used video recording in the interpretation of emotional expression displayed in the video sequence for the purpose of psychotherapy. The attribute sample shown in figure 2 above were extracted from experimental video sequence (Table 1) each attribute attraction correspond to a segment of the video sequence recorded at a given time, the time selected for extraction of attribute in the experiment was set to 0.01sec, any attribute which cannot be observe with the time frame will be set to "NILL" in the data set.

Table 1. The table shows the type of attributes derived from user corresponding to the time sequence. Those attributes were observed from the experiment (only partial segment of whole data set displayed here belonging to one user)

| F | G | H | I | J | K | L | M |
|---|---|---|---|---|---|---|---|
| Left Click Down | Left Click Release | Keypress | Real_Time | Mouse Movement | Cognitive_Sitimulis | HCO | USER_ID |
| NILL | LCR | NILL | 0.02 | NILL | 1 | NILL | 1 |
| LCD | NILL | NILL | 0.03 | Horizontal | GO | NILL | 1 |
| NILL | LCR | NILL | 0.05 | elliptical | PL | NILL | 1 |
| NILL | NILL | YES | 0.06 | Vertical | GO | NILL | 1 |
| LCD | NILL | NILL | 0.07 | NILL | GO | YES | 1 |
| NILL | LCR | NILL | 0.08 | elliptical | 1 | YES | 1 |
| LCD | NILL | NILL | 0.09 | elliptical | PI | YES | 1 |
| NILL | LCR | NILL | 0.1 | elliptical | PI | NILL | 1 |
| LCD | NILL | NILL | 0.11 | horizontal | GO | NILL | 1 |
| NILL | LCR | NILL | 0.12 | Horizontal | GO | NILL | 1 |
| LCD | NILL | NILL | 0.13 | Vertical | PI | NILL | 1 |

## 2.3 Bayesian networks for user identification

Bayesian network inference and classification utilities which are known as very robust against data sets which may contain substantial level of uncertainty. Bayesian Networks (BN) are known as "directed acyclic graphs" (*DAG*) which perform knowledge representation and reasoning even under uncertainty. They are also called directed Markov fields, belief networks, or causal probabilistic networks (Jensen, 1998). Bayesian networks are the probabilistic models which graphically encode and represent the conditional independence *(CI)* relationships among a set of data. In Bayesian networks each node represents a data attribute and is called a variable (it refers to a type of cognitive parameter in this work). The connections (arcs) between the nodes represent the dependency relationships of variables. Bayesian networks are very efficient tools for modelling the joint probability distributions of variables. In this system design initially the tools PowerConstructor© and PowerPredictor© are used (Cheng et al.,2002). The utilities accepts continuous variables and uses the Markov condition and conditional independence statements to build a network. The algorithm examines information regarding two related variables from a data set and decides if two variables are dependent. It also examines how close the relationship is between those variables. As far as the Bayesian theorem is concerned, it is a combination of likelihood and base rate which can be described by Bayesian formula as;

$$P(h|D) = \frac{p(h).p(D|h)}{p(D)} \qquad (1)$$

Here in Equation 1;
*p(h | D):* probability of h provided that evidence D is true ( h: hypothesis, D: evidence)
*p(D | h):* likelihood of *h* by belief,
*p(h) / p(D):* Base rate (number of total cases)

It was proven that (Hayes, 2000) human cognition always ignores the "Base rate" factor and it uses only the "likelihood" part of formula, which corresponds to belief. That is why Bayesian network is also called Belief network. The Bayesian approach has not only the advantage of including Base Rate factor, but also the definition of "uncertainty" concept within the classification process (Song and Yang, 2023). The Bayesian utility makes an examination of details of two closely related attributes (nodes) among the set of data available and then decides if the attributes are connected to each other or not indicating the relationship between attributes in the domain. The network topology is used to generate all possible independent relations between the attributes. A conditional mutual information of two nodes $Ai, Aj$ are investigated by the algorithm and make a decision regarding their relationship. This process can be denoted by the formula ;

$$I(Ai, Aj \backslash C) = \sum_{ai,aj,c} P(ai, aj, c) \log \frac{P(ai,aj \backslash c)}{P(ai \backslash c)P(ai \backslash c)} \qquad (2)$$

Wher*e*; **t** : the threshold,
**C :** a set of nodes
**c** : vector representing one instance of attribute C,

(Orun Seker, 2012). In the condition where $I(Ai, Aj \backslash C)$ is smaller than the threshold t, then $(Ai \text{ and } Aj)$ are conditionally independent while the value of $P(ai, aj, c)$ can be extracted from conditional probability table.

# 3 Results and discussion

Within the proposed system's experiments for remote identification of individual's cognitive characteristics, PC keyboard, PC mouse, graphical GUI domain (stimulant) and mouse/keyboard monitoring utility have been used. This is at some extend the simulation of "real" cyber security environment in which some other remote stimuli techniques will be used such as pop-up screens, social media domains, adverts, etc. In this case, to construct the data set, personal cognitive behavioural attributes have been specified having link to Graphical test domain and PS keyboard/mouse output values in a time sequence (Figure 1). It should be noted that, Similar techniques had already been used in the previous work where the game monitor was used as a stimuli and cognitive properties were collected by camera and skin sensors (Orun and Seker, 2012). Those proposed game-based system had already identified the individuals at high accuracy. But this work is specifically based on keyboard/mouse signals which is more suitable for remote identification of the cognitive factors. As show in Figure 1, the user is stimulated by the graphical domain, to exhibit his/her personal cognitive behaviours via keyboard/mouse usage pattern which refers to his/her personal signature. The arrows symbolized the typical mouse motions on the dynamic test objects (Irigbhogbe,2015) where the user follows a particular path specific to his/her personal cognitive choice. Such graphical domains simply stimulate the user's optional behaviours by telling him/her to click on certain graphical object by mouse motion and type the correct word on them via keyboard. Those behaviours are meanwhile recorded by the mouse/keyboard monitoring utility simultaneously (Figure 1; right hand side). The time-sequence data are used to extract attributes to build ultimate data set, whose attributes are listed as below. The attributes are extracted form the mouse/keyboard motions and output signals whose cases refer to each row of the data set.

## 3.1 Graphical user environment for cognitive data collection

The custom designed graphical user environments are prepared to be used as cognitive stimuli which will lead to cognitive data collection as were already exploited in some of the early studies (Bostan, 2010; Orun et al., 2016). The design of graphical characteristics are randomly selected and arranged not in particular way. This is because the data generated by the user behaviour would include sufficient information for user's cognitive characteristics to be unveil by the Artificial intelligence (AI) methods like Bayesian networks (Orun and Seker, 2012)

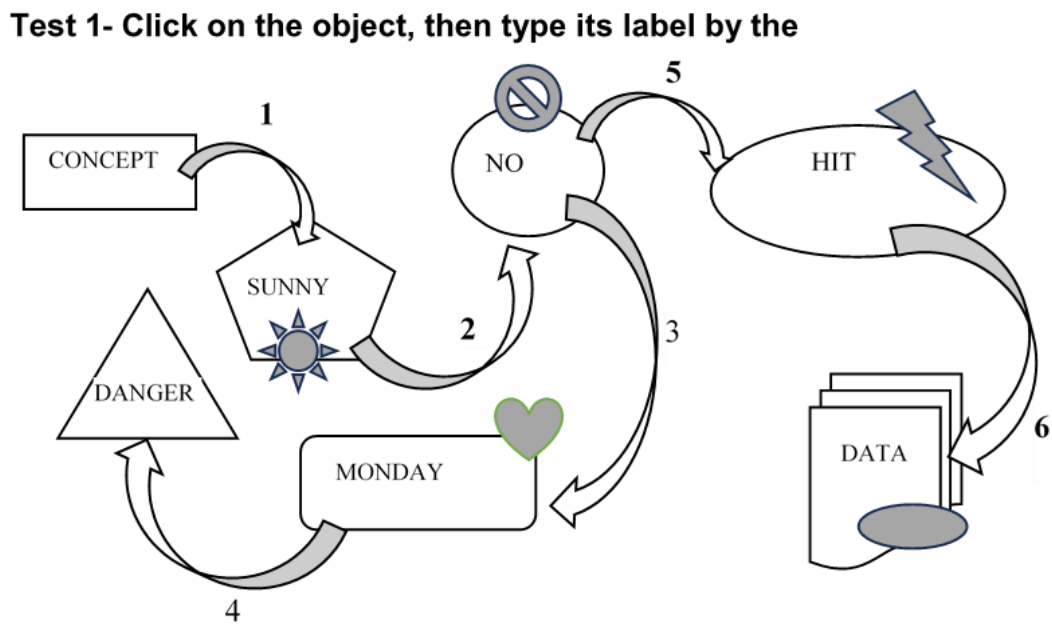

Figure 1. Graphical stimuli domain test environments are used to stimulate user to exhibit his/her cognitive characteristic behaviours. Which then help construct data set to be analysed by Bayesian network classifier for remote identification of an individual. The arrows symbolized the typical mouse motions on the dynamic test objects (Irigbhogbe,2015) where the user follows a particular path specific to his/her personal cognitive choice. The mouse motion / location coordinates and keyboard presses are shown on the numerical table (Table2)

By the remote identification tests, the identity Classification accuracy between the individual 1 and 2 have been obtained as 79% in first test, and 94% in second test, which is the maximum accuracy obtainable with the available facilities, data set used. This accuracy rate naturally would decrease by the increase of individuals' population, but may increase with better and more precise data acquisition at higher volume. In data set of keyboard/mouse outputs, each row correspond to a case at certain Signal acquisition time (e.g. every 0.1 second) and the columns refer to attribute values. The data set are divided by two parts by the ratios of 3/2 approximately. One part is used for the training session of Bayesian network, and the other half of data is used for tests of classification stage. The data set contains the attributes labelled as;

- o *Left Click Down*: generated by the left mouse button as is kept press down
- o *Left Click release*: generated by the left mouse button as is released
- o *Keypress:* Keyboard activation signal
- o *Real time:* time recordings of all activity signals generated by keyboard or mouse
- o *Mouse movement*: signals generated by mouse motion (e.g. x, y, horizontal, elliptical, vertical
- o or no motion)
- o *Cognitive stimuli:* user's option of graphical object selection on the test screen.

The networks' specifically selected parameters for their classification optimisations and their accuracy results are shown in Table 2 for Test1 and Test 2. The links between user ID and Cognitive Stimuli attributes is also shown in Figure 3 in the Bayesian Network configurations. In Figure 3, the classification test 1 and 2 where identity of individuals can be distinguished and identified by the networks class node labelled as "user_ID". As it seems the classification process is directly achieved by the nodes;"MouseY", "MouseX", "time-metrics",and indirectly made by "left mouse click" and "keyboard".

Table 2. The user's behaviours attributes as recorded in the sequential time domain. Such as mouse coordinates, mouse click or release, keyboard usage and time elapsed between them.

| *Case No* | *Mouse motion X* | *Mouse motion Y* | *Time between events* | *Mouse/Keyboard Usage* | *Case No* | *Mouse motion X* | *Mouse motion Y* | *Time between events* | *Mouse/Keyboard Usage* |
|---|---|---|---|---|---|---|---|---|---|
| 9 | 372 | 659 | 111 | Left click release | 21 | 452 | 530 | 78 | Left click release |
| 10 | 506 | 299 | 3562 | Left click down | 22 | 452 | 530 | 2360 | Keypress r |
| 11 | 506 | 299 | 214 | Left click release | 23 | 452 | 530 | 937 | Keypress space |
| 12 | 430 | 329 | 197 | Left click down | 24 | 452 | 530 | 156 | Keypress a |
| 13 | 430 | 329 | 197 | Left click release | 25 | 452 | 530 | 1110 | Keypress r |
| 14 | 602 | 425 | 1844 | Left click down | 26 | 452 | 530 | 800 | Keypress o |
| 15 | 602 | 425 | 94 | Left click release | 27 | 452 | 530 | 308 | Keypress n |
| 16 | 601 | 444 | 520 | Left click down | 28 | 452 | 530 | 2693 | Keypress tab |
| 17 | 601 | 444 | 78 | Left click release | 29 | 470 | 588 | 4765 | Left click down |
| 18 | 608 | 471 | 640 | Left click down | 30 | 470 | 588 | 125 | Left click release |

The specific Bayesian utility PowerConstructor$^{TM}$ allows parametric optimisation to maximize the accuracy of classification results obtained. The parametric options are listed in Table 3 for Test1 and Test2.

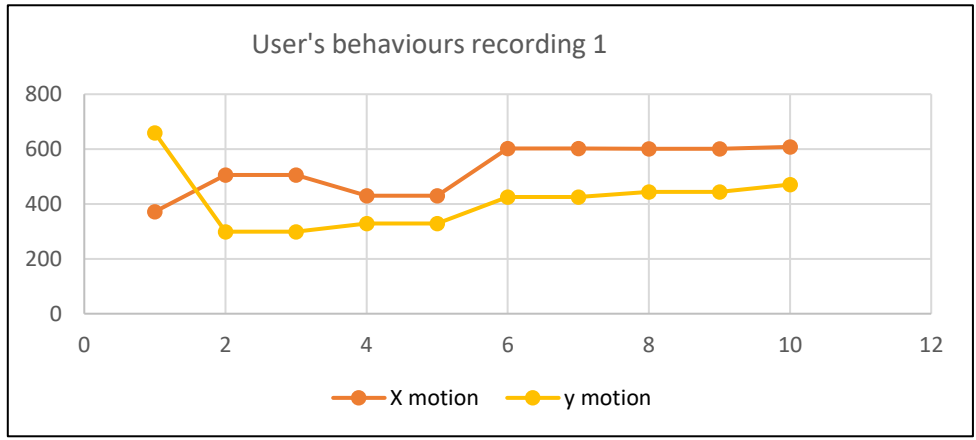

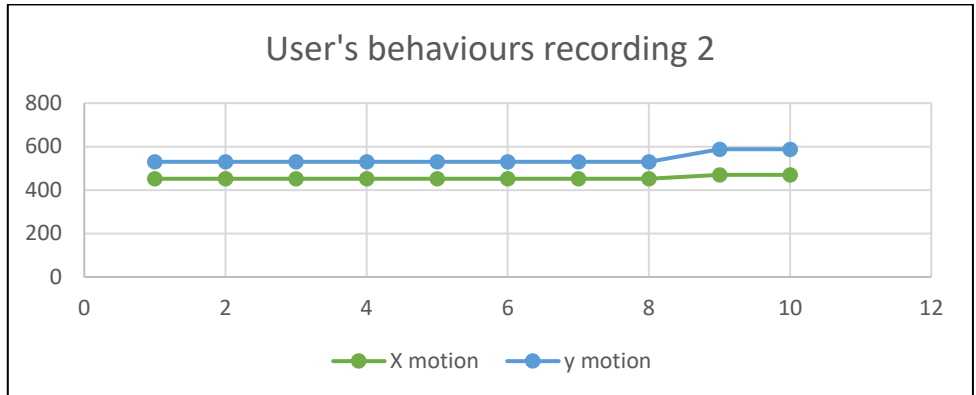

Figure 2. User's behavioural changes displayed by mouse x and y motions as two separate attributes in a time sequence (comparative time) as the graphs extracted from Table 2 and based on Graphical stimuli domain in Figure 1. The graph representations aim to describe typical user attributes changes in time sequences domain but not sufficient to provide substantial discrimination power without any AI method's contribution.

Table 3. The Bayesian Network parameters selected for Test 1 and 2 for their classification optimization

| Test 1 | Test 2 |
| --- | --- |
| discretisation method = equal frequency | discretisation method = equal width |
| Number of intervals = 5 | Number of intervals = 10 |
| Attributes Connection threshold t = 0.1 | Attributes Connection threshold t = 0.1 |
| Classification Accuracy = 79% | Classification Accuracy = 94% |

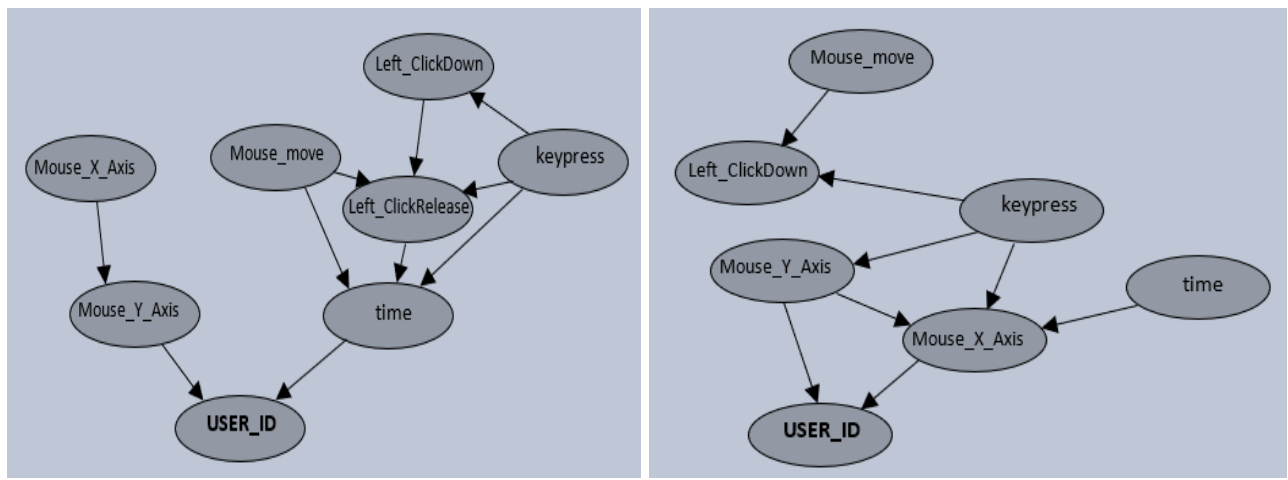

Figure 3. The Bayesian Network classification test 1 and 2 (on the left and right respectively) where identity of individuals are distinguished and identified (classified) by the class node labelled as "user_ID". The classification is directly made by the nodes; "MouseY", "MouseX", "time-metrics", and indirectly made by "left mouse click" and "keyboard" (Irigbhogbe, 2015)
The Bayesian Network parameters selected for Test 1 and 2 for their classification optimization are shown in Table 3.

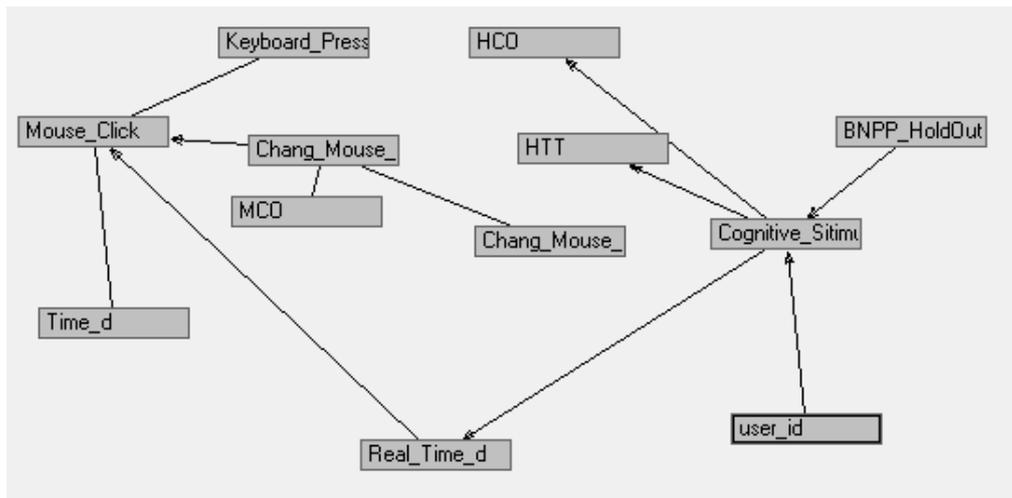

Figure 4. The effective role of "cognitive stimuli" attribute on the process for classification of user's identity is demonstrated by the link which is automatically established by Bayesian inference tool PowerConstructor$^{TM}$. In the network; Cognitive stimuli refer to features like number, icon, button, text, etc. MCO, HCO and HTT refer to mouse motion attributes (motion style: elliptical, horizontal, etc.) which reflect to cognitive behaviours of the user.

# 4 Conclusion

The proposed "cognitive AI" based sub-system is scheduled to be a part of an ultimate more comprehensive Cyber-security system and would have an open-ended design being appropriate for a further evaluation such as utilizing other AI techniques like deep learning, better graphical environments, and user behaviours sensing and recording utilities, etc. It should be noted that, the proposed system with the limited facilities here is at some extend the simulation of "real" cyber security environment in which some other more realistic and pragmatic remote stimuli techniques would also be used such as pop-up screens, social media domains, adverts, etc. to stimulate the intruders or hackers. Here the system flexibility will allow an unlimited number of users falling into two or three major categories (e.g. as intruder, suspicious or innocent) and each would have specific cognitive characteristic pattern. The system design will also be flexible to adapt with new fashion communication and online shopping instruments (e.g. tablets, smart phones, etc.) which are potentially prone to cyber-attacks. Within this work the tests yields 94% classification accuracy to distinguish between user1 and user2 correctly, this result shows the feasibility of developing a real-world remote attack detection system based on the analysis techniques on this paper. It is understood that authentication system are capable of producing a TRUE or FALSE output i.e. if a system is able to understand the cognitive behaviour of its current user, it can tell if it is an attacker or a legitimate user by comparing a previously stored profile of the legitimate user to the current user. To defend again attack, any suspicious manipulation of the system GUI not corresponding with a previously register profile will have to be authenticated or identified.